\documentclass[aps,pre,twocolumn,amsfonts,amssymb,superscriptaddress,showpacs]{revtex4}
\usepackage{placeins}
\usepackage{graphicx}
\usepackage{amsmath}
\usepackage{times}
\usepackage{color}

\begin{document}
\title{Stochasticity and evolutionary stability}
\author{Arne Traulsen}
\affiliation{Program for Evolutionary Dynamics, Harvard University, Cambridge MA 02138, USA}
\author{Jorge M.\ Pacheco}
\affiliation{Program for Evolutionary Dynamics, Harvard University, Cambridge MA 02138, USA}
\affiliation{Centro de F{\'\i}sica Te{\'o}rica e Computacional, 
             Departamento de F{\'\i}sica da Faculdade de Ci{\^e}ncias, 
             P-1649-003 Lisboa Codex, Portugal}
\author{Lorens A. Imhof}
\affiliation{Program for Evolutionary Dynamics, Harvard University, Cambridge MA 02138, USA}
\affiliation{Statistische Abteilung, Universit{\"a}t Bonn, Adenauerallee 24-42, 53113 Bonn, Germany}
\date{ 4 August 2006, Physical Review E 74, 021905}
\begin{abstract}
In stochastic dynamical systems, different concepts of stability can be obtained in different 
limits. A particularly interesting example is evolutionary game theory, which is traditionally 
based on infinite populations, where strict Nash equilibria correspond to stable fixed points 
that are always evolutionarily stable. However, in finite populations stochastic effects can drive the system away from strict Nash equilibria, which gives rise to a new concept for evolutionary stability. The conventional and the new stability concepts may apparently contradict each other leading to conflicting predictions in large yet finite populations.
We show that the two concepts can be derived from the frequency dependent Moran process in different limits. Our results help to determine the appropriate stability concept in large finite populations. 
The general validity of our findings is demonstrated showing that the same results are valid employing 
vastly different co-evolutionary processes. 
\end{abstract}
\maketitle

\section{Introduction}

Evolutionary game theory \cite{maynard-smith:1982to} provides a powerful and unifying mathematical framework widely used in scientific areas as diverse as biology \cite{nowak:2004aa}, economics \cite{gintis:2000bv}, and social sciences. 
More recently, it has been attracting growing interest in physics  \cite{ebel:2002aa,szabo:2002te,santos:2005pm}, as several sophisticated techniques developed in the physics of complex systems have provided useful insights into this interdisciplinary framework \cite{hauert:2005mm}. Originally, evolutionary game theory was formulated in terms of infinite populations and the corresponding replicator dynamics \cite{taylor:1978wv,hofbauer:1998mm}. As a nonlinear dynamical system, these equations are of great interest being formally equivalent to the well studied Lotka-Volterra equations \cite{hofbauer:1998mm} . Much of our present intuition is based upon this deterministic framework that analyzes nonlinear systems of ordinary differential equations.  However, any real population has finite size and also individual-based computer simulations in unstructured or structured populations \cite{szabo:1998wv,holme:2003mm,traulsen:2004aa,vukov:2005fa,santos:2005bb,rauch:2006aa,santos:2006pn} always deal with finite populations. As is well known in physics, the behavior of a finite system can depart significantly from its infinite counterpart. 

Recently, the concept of evolutionary stability has been specifically investigated for finite populations \cite{nowak:2004pw,wild:2004aa} and the traditional concept of evolutionary stability has been challenged. Analyzing the connection between the two frameworks of evolutionary stability we show that the difference between them is not solely based on finite size effects. 

When selection is frequency independent, the idea of evolutionary stability is simple: A population evolves until a stable fixed point is reached at which fitness is maximized. In contrast with this situation, in evolutionary game theory the fitness of an individual depends on the type and frequency of its competitors. Hence, the optimization of individual fitness can even lead to the decline of the average fitness, as in the Prisoner's Dilemma where each individual is better off not cooperating, although mutual cooperation leads to higher fitness \cite{axelrod:1984yo}.

The standard definition of evolutionary stability formulated for infinite populations 
\cite{maynard-smith:1982to} is equivalent to strong stability of the corresponding fixed 
point in the replicator dynamics \cite{hofbauer:1998mm}: A strategy is an evolutionarily 
stable strategy (ESS) if individuals with this strategy are always better off than a small 
fraction of mutants. A similar result holds for infinite populations  
subject to external noise: In the stochastic replicator dynamics of Fudenberg and
Harris \cite{fudenberg:1992bv},
the population stays nearly all the time close to an ESS if the ESS corresponds
to an interior state or if the payoff matrix satisfies a certain definiteness condition
\cite{imhof:2005aa}. Moreover, in the stochastic replicator dynamics, 
strict Nash equilibria are always asymptotically stochastically stable, provided the 
impact of the external noise is not too strong.
However, for finite populations the concept of evolutionary stability has been challenged, as an ESS defined as above
is not necessarily stable anymore whenever the population size is finite \cite{nowak:2004pw}. Nowak {\em et al.} 
introduced a concept for evolutionary stability in finite populations (ESS$_{\rm N}$) 
proposing the additional requirement that selection has to oppose replacement by other 
strategies due to random drift \cite{nowak:2004pw}. For coordination games corresponding to 
bistable situations, they also derived the condition under which selection favors the 
replacement of one strict Nash equilibrium by the other. They found that replacement occurs if the unstable fixed point that is always present in coordination games is closer than $1/3$ to the strategy to be replaced. 
This finding obtained for weak selection contradicts the idea of evolutionary stability in infinite populations, where such a replacement cannot occur. Furthermore, even in simulations of finite populations such a replacement has not been demonstrated yet, as selection is usually strong in these systems. Here we show that these two apparently contradicting concepts actually emerge as different limiting results of a unified treatment.
Both the traditional ESS concept and the ESS$_{\rm N}$ concept can be derived from the frequency dependent Moran process in different limits. However, these results do not rely on Moran dynamics. Indeed, their general validity is shown to apply to other types of co-evolutionary dynamics, such as the frequency dependent Wright-Fisher process  \cite{imhof:2006aa} and the local update process investigated in \cite{traulsen:2005hp}.
 
 The remainder of this paper is organized as follows: In Sec.~\ref{moransec}, we introduce the frequency dependent Moran process and discuss how different limits lead to different concepts of evolutionary stability. In Secs.\ \ref{lursec} and \ref{wfsec}, we show how the results transfer to the Local update process and the frequency dependent Wright-Fisher process. Finally, we discuss implications on the fixation times in Sec.\ \ref{timesec}.
 
 \section{The frequency dependent Moran process}
 \label{moransec}
 
 In evolutionary game theory, the fitness of an individual is determined by the payoff from its interactions with others.  An $A$ individual will obtain a payoff $a$ from an interaction with $A$ individuals and $b$ from $B$ individuals. Similarly, $B$ individuals gain $c$ from $A$ and $d$ from $B$. Hence, the payoffs are given by 
\begin{eqnarray}
\pi_A & = & a \frac{i-1}{N-1} +b \frac{N-i}{N-1}\ \\ 
\pi_B & = & c \frac{i}{N-1} +d \frac{N-i-1}{N-1}
\end{eqnarray}
for the types $A$ and $B$, respectively. $N$ is the population size and $i$ is the number of $A$ individuals in the population. Self-interactions have been excluded. 
For the frequency dependent Moran process \cite{nowak:2004pw}, the fitness of an individual is a linear combination of a background fitness and the payoff from interactions, $f_A=1-w+w \pi_A$ for $A$ individuals and analogously for $B$. The parameter $0\leq w \leq 1$ measures the intensity of selection. For $w \ll 1$, selection is weak and the payoff of the game has only a marginal influence, whereas the background fitness becomes negligible for $w \to 1$. 

We assume that the fitness is always positive. 
An individual chosen proportional to its fitness produces one identical offspring which replaces a randomly chosen individual. 
The probability to increase the number of $A$ individuals from $i$ to $i+1$ is
\begin{equation}
T^{+}(i) = \frac{i\,f_A}{i\, f_A + (N-i) f_B} \frac{N-i}{N},
\end{equation}
whereas the probability to decrease it from $i$ to $i-1$ is 
\begin{equation}
T^{-}(i) = \frac{(N-i)f_B}{i f_A + (N-i) f_B} \frac{i}{N}. 
\end{equation}
Since $T^{-}(N)=0$ and $T^{+}(0)=0$, this process has absorbing states at $i=0$ and $i=N$. For large $N$, the Master equation describing this process can be approximated by a Fokker-Planck equation 
with drift $a(x) \approx T^{+}(x)-T^{-}(x)$ and diffusion $b^2(x)\approx(T^{+}(x)+T^{-}(x))/N$, where $x=i/N$ is the fraction of type $A$ in the population \cite{traulsen:2005hp}.
This Fokker-Planck equation corresponds to the stochastic differential equation
\begin{equation}
dx = a(x)dt + b(x) dW(t),
\label{langevin}
\end{equation}
where $W(t)$ is the Wiener process, $\langle W(s) W(t) \rangle = \min(s,t)$ \cite{kampen:1997xg}. 
For the frequency dependent Moran process, we have
\begin{eqnarray}
a(x) &\approx&  x(1-x) \frac{ f_A(x)-f_B(x) }{x f_A(x)+(1-x)f_B(x)} \\
b^2(x) &\approx& {\frac{x(1-x)}{N} \frac{f_A(x)+f_B(x)}{x f_A(x)+(1-x)f_B(x)}}. 
\end{eqnarray}
For fixed selection intensity $w$ and $N \to \infty$, $b(x)$ vanishes and a deterministic replicator equation is obtained from the frequency dependent Moran process \cite{traulsen:2005hp}. 
In this deterministic limit, a stable coexistence of $A$ and $B$ is possible.

However, if $N w \ll 1$, the stochasticity is retained even in the limit of infinite population size and the system will eventually get absorbed in the state $x=0$ or $x=1$. The probability that the system gets absorbed in $x=1$ starting from $x_0$ (in other words, the probability to reach the state with $A$ individuals only) can be computed from the drift and diffusion term as 
\begin{equation}
\phi(x_0) = \frac{S(x_0)}{S(1)} 
\hspace{0.5cm}{\rm where} \hspace{0.3cm}
S(x)=\int_0^x e^{-\int_0^y \Gamma(z) dz } dy,
\label{fixation}
\end{equation}
and $\Gamma(z)=2 a(z)/b^2(z)$ \cite{ewens:1979qe}.
It is noteworthy that there are similarities between our absorption probabilities Eq.~(\ref{fixation}) and the results of Ref.~\cite{laessig:2002aa} in which a quantum description of game dynamics has been explored.

For neutral selection, $w=0$, the drift term vanishes, $a(z)=0$, and the fixation probability is simply the initial fraction of type $A$, $\phi(x_0)=x_0$. 
For $w \ll 1$, the ratio $\Gamma(z)$ that determines the fixation properties of the process becomes
\begin{equation}
\label{gammamoran}
\Gamma(z) \approx N w ( \pi_A -\pi_B) = 
N w \left( \alpha z +\beta \right),
\end{equation}
where $\alpha = (a-b-c+d) N/(N-1)$ and $\beta = (-a+bN-dN+d)/(N-1)$. 
A particular interesting case is $N w \ll 1$, in which the fixation probability 
Eq.\ (\ref{fixation}) is given by  
\begin{equation}
\phi(x) \approx x+ \frac{w}{6} N x (1-x) \left(\alpha (1+x) +3 \beta \right).
\label{moranfixation}
\end{equation}
For $x=1/N$, this reduces to the fixation probability of a single mutant in the limit of weak selection, which has been computed directly from the Moran process in \cite{nowak:2004pw}. 
Selection favors $A$ replacing $B$ if the fixation probability of a single $A$ mutant 
is higher than the fixation probability of a neutral mutant, which amounts to $\phi(x)>x$.
For the evolutionary stability of the state with $B$ individuals only, the condition $\phi(x)>x$ is of interest in the vicinity of $x=0$. After derivation with respect to $x$ we find for $x=0$ the condition $S(1)<S'(0)=1$. For $N \to \infty$ and $w \to 0$ with $N w \ll 1$, the development of $S(1)<1$ reduces to 
$a+2b>c+2d$. 

For $a>c$ and $b>d$, the fitness of $A$ is larger than the fitness of $B$ for any number of mutants. In this case, the condition $S(1)<1$ is always fulfilled and $A$ will invade the population with a higher probability than a neutral mutant. On the other hand, for $a<c$ and $b<d$, the strategy $A$ is disadvantageous and the probability for its invasion is smaller than that of a neutral mutant. 
However, the most interesting case is associated with those situations in which $\pi_A-\pi_B$ changes sign for a certain $x$. Here, we concentrate on coordination games in which $a>c$ and $b<d$. 
In these games, the situation with 100\% $A$ or 100\% $B$ individuals are strict Nash equilibria: If the population is dominated by type $A$, a $B$ mutant is disadvantageous, but if $B$ dominates, $A$ individuals become disadvantageous. Coordination games are bistable, i.e.\ for an intermediate number of mutants the preferred direction of the evolutionary process changes, as the drift term $a(x)$ changes sign. 
Whereas in the deterministic replicator equation $x=0$ and $x=1$ are stable fixed points, stochasticity can lead from one point to the other, similar to a physical process in which stochastic fluctuations enable a particle to overcome an energy barrier. 
In asymmetric situations, stochasticity favors one equilibrium over the other, depending on the position of the unstable fixed point $x^{\ast}$ between them. 
When this point is close to $0$, it is relatively easy for $A$ to invade, as a small group of $A$ individuals changes the sign of the drift term $a(x)$.
More specifically, for $a>c$ and $b<d$ selection favors $A$ replacing $B$ if
\begin{equation}
 x^{\ast} =  \frac{d-b}{a-b-c+d}  < \frac{1}{3} 
\label{onethirdrule}
\end{equation}
where $x^{\ast}$ is the unstable equilibrium of the game at which the fitness of both strategies is the same, $\pi_A = \pi_B$. 
This result coined as $1/3$-rule by Nowak {\em et al.} can be derived directly from the exact fixation probabilities of the Markov chain and is valid for $N w \ll 1$.
It is a central result of the ESS$_{\rm N}$ concept \cite{nowak:2004pw}.
It can also be obtained in the same limit from Eq.\ (\ref{moranfixation}) with $x=1/N$. 
Furthermore, Eq.\ (\ref{moranfixation}) shows that this $1/3$-rule extends beyond the case of a single mutant and is also obtained if the fixation probability of a fixed number of mutants is investigated.

\begin{figure}[hpbt]
\begin{center}
{\includegraphics[width=9.0cm]{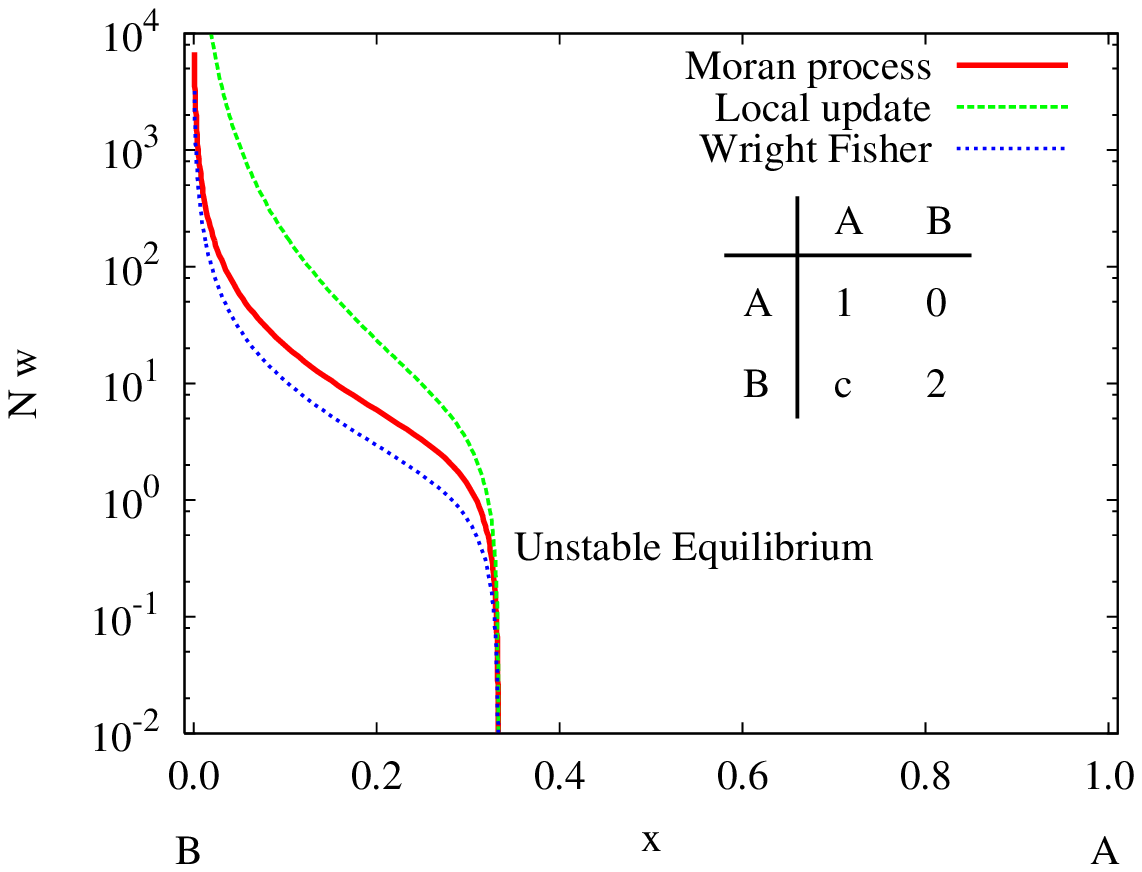}}
\end{center}
\vspace{-0.6cm}
\begin{flushright}
{\includegraphics[width=7.0cm]{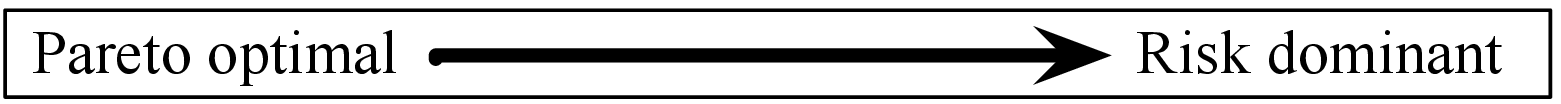}}
\end{flushright}
\vspace{-0.8cm}
\begin{center}
\caption{
(color online) The favored equilibrium in a bistable situation given by a coordination game depends on 
the position of the unstable equilibrium $x^{\ast}$. Here, the position of $x^{\ast}$ for 
which selection in finite populations favors the replacement of the Pareto optimal equilibrium $B$ (with the higher payoff) by the risk dominant strategy $A$ (with a larger basin of attraction) is computed numerically for different $N$ and fixed $w = 0.001 $.  $x^{\ast}$ is shifting towards the pure $B$ strategy with increasing $N w$, i.e.\ decreasing degree of stochasticity. For $N w \ll 1$, the equilibrium has to be at $x^{\ast} \leq 1/3$ for $A$ to replace $B$ irrespective of the process at stake (the local update process, the Moran process, or the frequency dependent Wright-Fisher process). For $N w \to \infty$ the equilibrium converges to the pure $B$ strategy, as expected for the replicator dynamics where selection never favors $A$ replacing $B$ if $B$ is a stable  fixed point. Although these limiting results hold for all three processes, the position of $x^{\ast}$ will depend on the specific co-evolutionary process and the payoff matrix
($a=1$, $b=0$, $c=3-2/x^{\ast}<-3$, $d=2$, $w=0.001$).
\label{onethirdfig}
}
\end{center}
\end{figure}

In general, the stochasticity arising from finite populations is qualitatively different from the Gaussian noise incorporated in Eq.\ (\ref{langevin}). Therefore, it is remarkable that the approximation of population dynamics in finite populations with stochastic differential equations agree for weak selection perfectly with the direct calculation for arbitrary $N$, although the Langevin approximation implicitly assumes large populations. 
However, if $w$ remains bounded away from zero, we can no longer assume $N w \ll 1$ for $N  \to \infty$ and the $1/3$-rule is violated, as illustrated in Fig.\ \ref{onethirdfig}. 
In this case, we obtain for $w \ll 1$
\begin{eqnarray}
\label{moranonethird}
S(1) & \approx & \int_0^1 \exp \left[-N w \int_0^y  \left( \alpha x +\beta \right)dx\right]dy  \\ \nonumber
& = & 
\sqrt{\frac{\pi}{2 \alpha} \frac{1}{N w}} e^{z_0^2 } 
\left[ {\rm erf} \left( z_1 \right) - {\rm erf} \left( z_0 \right) \right], 
\end{eqnarray}
where ${\rm{erf}}\,(x)=\frac{2}{\sqrt{\pi}}\int_0^x dy\, e^{-y^2}$ is the error function
and $z_i = (\alpha i + \beta) \sqrt{N w/(2 \alpha)}$.
For coordination games with $a>c$ and $b<d$,  the right hand side of Eq.\ (\ref{moranonethird}) diverges for $N \to \infty$ and fixed $w \ll 1$. Therefore, $S(1)>1$ and selection {\it never} favors $A$ replacing $B$ in this setting. As $B$ and $A$ are ESS in coordination games, this is exactly what the traditional evolutionary stability predicts. Hence, we have shown that evolutionary stability depends on the quantity $N w$, i.e. the size of the population times the intensity of selection. $N w \ll 1$ leads to the ESS$_{\rm N}$ concept from \cite{nowak:2004pw}, whereas $N w \gg 1$ reveals the traditional ESS concept \cite{maynard-smith:1982to}. There is no contradiction between these notions of evolutionary stability: 
They both constitute two extreme limits of the quantity $N w$. Indeed, as explicitly shown in the next two Sections, this result does not depend on the frequency dependent Moran process adopted so far.
 
For intermediate values of $N w$, the position of the unstable equilibrium $x^{\ast}$ such that selection favors the replacement of $B$ by $A$ can be solved numerically from $S(1)=1$ for a given payoff matrix, as shown in Fig.~\ref{onethirdfig}. 
For $x<x^{\ast}$ selection favors replacement of $B$ by $A$, for $x>x^{\ast}$ this replacement is not favored.   
Eq.\ (\ref{moranonethird}) depends on the payoffs as a function of $\alpha$ and $\beta$ and not only as function of $x^{\ast}$. Therefore, there is no universal relation analog to the $1/3$-rule for fixed $N w$. In general, the position of the unstable equilibrium will depend on all entries in the payoff matrix and not only on a certain combination of them.
In Fig.~\ref{onethirdfig}, one parameter of the payoff matrix is varied and the resulting position of $x^{\ast}$ is shown. 
For coordination games ($a>c$ and $d>b$) with $d>a$, $B$ is a Pareto optimal equilibrium that maximizes payoffs, as the equilibrium with $A$ players only yields a lower payoff. Furthermore, the condition $a-c>d-b$ implies that $A$ is risk dominant, i.e.\ equilibrium $A$ has a larger basin of attraction than $B$. 
If the level of stochasticity is small, $N w \gg 1$, replacement of the Pareto optimal equilibrium only occurs if the basin of attraction of this equilibrium is very small. For high stochasticity, $N w \ll 1$, replacement of the Pareto optimal by the Risk dominant equilibrium is much more likely. 

We note that we do not consider a thermodynamical limit here, as the properties of the system are not conserved. A more detailed account on the thermodynamical limit of the frequency dependent Moran process is given in \cite{chalub:2006cc}.
 
\section{The Local update processes}
 \label{lursec}

To demonstrate the general validity of our result for evolutionary stability, we show that the two limits $Nw \ll 1$ and $N w \gg 1$ lead to the same result for two vastly different processes. In this Section, we consider the local update process discussed in \cite{traulsen:2005hp}. In this process, two individuals are chosen at random from the population. Only the choice of two different individuals can change the composition of the population. In mixed pairs, the $A$ individual replaces the $B$ individual with probability 
\begin{equation}
p = \frac{1}{2}+ \frac{w}{2} \frac{\pi_A-\pi_B}{\Delta \pi}, 
\end{equation}
where $\Delta \pi$ is the maximum possible payoff difference.  
The $B$ individual replaces the $A$ individual with probability $1-p$.
As for the Moran process, $w$ determines the intensity of selection: For small $w$, stochastic effects are frequent, whereas $w \to 1$ leads to strong selection. 
This process is described by the transition probabilities 
\begin{equation}
T^{\pm}(i) = \frac{i}{N} \frac{N-i}{N} \left(\frac{1}{2}\pm \frac{w}{2} \frac{\pi_A-\pi_B}{\Delta \pi} \right).  
\end{equation}
Again, the birth-death process can be approximated by a stochastic differential equation, where the drift $a(x)$ and the diffusion $b(x)$ are given by \cite{traulsen:2005hp}
\begin{eqnarray}
a(x) &\approx&  x(1-x) \frac{w}{\Delta \pi}  \left[\pi_A(x)-\pi_B(x) \right] \\
b^2(x) &\approx& {{x(1-x)}/{N} }. 
\end{eqnarray}
In this case, 
$\Gamma(x) = 2 N w  (\alpha x +  \beta)/\Delta \pi$ ,
which  is linear in $x$ for arbitrary intensity of selection $w$. Hence, we can compute the fixation probability Eq.\ (\ref{fixation}) explicitly, 
\begin{equation}
\phi(x) \approx
\frac{{\rm erf}\left[  \zeta(x)\right]-
{\rm erf}\left[  \zeta(0) \right]
}{{\rm erf}\left[  \zeta(1) \right]-
{\rm erf}\left[ \zeta(0)\right]
},
\label{fixproblocal}
\end{equation}
where $\zeta(x) = (\alpha x + \beta) \sqrt{N w /(\alpha \Delta \pi)}$.
For weak selection $w \ll 1$, the expansion of the error functions in Eq.\ (\ref{fixproblocal})
leads to Eq.\ (\ref{moranfixation}) again with $w \to 2 w /\Delta \pi$. 
For $x=1/N$, this reduces exactly to the result derived directly from the 
Markov chain in \cite{traulsen:2005hp}. 
Again, we can ask under which circumstances selection favors $A$ replacing $B$. 
Since $\Gamma(x)$ is up to a constant factor identical to the corresponding
function for the Moran process for weak selection,  
Eq.\ (\ref{moranonethird}) is again obtained for the local update process.
However, here it is valid for {\em any} selection intensity $w$. 

For $N w \ll 1$,
Eq. (\ref{onethirdrule}) is recovered again. Hence, the $1/3$-rule is valid for weak selection even for
the local update process. 

However, for $N w \gg 1$ 
selection will never favor $A$ replacing $B$ in a coordination game as $\phi(x)<x$ for $x \ll 1$, see also Fig.\ \ref{onethirdfig}. This implies that any finite $w$ guarantees evolutionary stability in the limit $N \to \infty$. Hence, $N w \gg 1$ leads back to the traditional concept of evolutionary stability as obtained above for the Moran process. 

\section{The frequency dependent Wright Fisher process}
 \label{wfsec}
 
As a third example, we consider the frequency dependent Wright-Fisher process, in which individuals reproduce proportional to their fitness $f_A = 1-w+w \pi_A$, similar to the Moran process. However, here reproduction is not directly connected to death. Instead, the new generation is sampled at random from a large population constituted by the offspring of all individuals. Hence, each time step corresponds to an entire generation with $N$ time steps in the Moran process. The dynamics of this process is given by the transition matrix \cite{imhof:2006aa}
\begin{equation}
T_{i \to j} = \left( \!\! \begin{array}{c} N \\ j \end{array} \!\! \right) \!\!
\left(\frac{i f_A}{i f_A +(N\!-\! i) f_B} \right)^{\! j} \!\!
\left(\frac{(N\!-\! i) f_B}{i f_A +(N\!-\! i) f_B} \right)^{ \! N-j}
\nonumber
\end{equation}
resulting from binomial sampling. As this matrix is not tri-diagonal, we explicitly derive $a(x)$ and $b(x)$. The drift term is given by
$a(x) =  \langle x_{t+\Delta t} - x_t \rangle/ \Delta t,$
where $\Delta t$ is the time step and $x_t$ is the fraction of $A$ individuals at
time $t$, $x_t = i_t/N$. With $\Delta t = 1/N$, this yields
\begin{equation}
a(x) \approx  N  x(1-x) \frac{f_A(x)-f_B(x)}{x f_A(x)+(1-x)f_B(x)}.
\end{equation}
Similarly, the diffusion $b^2(x) = \langle \left( x_{t+\Delta t} - x_t \right)^2 \rangle/\Delta t$
is 
\begin{eqnarray}
b^2(x) & \approx & {x(1-x)} \\ \nonumber
& \times &\frac{{f_A(x)\,f_B(x)+ N x (1-x) (f_A(x)-f_B(x))^2}}{\left(x f_A(x)+(1-x)f_B(x) \right)^2}.
\end{eqnarray}

Again, we first consider weak selection. 
The population size enters in the diffusion term as $N w^2$. Hence, for weak selection $N w^2 \ll 1$ is required (in contrast to the processes discussed above, where $N w \ll 1$ was sufficient).
For weak selection, the diffusion term becomes $b(x) = \sqrt{x(1-x)}$ whereas the drift term is
$a(x) = N  x(1-x) \left(f_A(x)-f_B(x) \right) $. 
Hence, 
$\Gamma(x)\approx 2 N w  \left(\pi_A-\pi_B \right)$,
which is up to a factor $2$ identical to the corresponding equation
for the Moran process, cf.\ Eq.\ (\ref{gammamoran}).
Consequently, all results for {\em weak} selection, transfer to the frequency dependent Wright-Fisher process. In particular, Eqs. (\ref{moranfixation}) and (\ref{onethirdrule}) are obtained again, i.e.\ the 1/3- rule is fulfilled, as shown by a different approach in \cite{imhof:2006aa}. 

For $w \ll 1$ and 
$N w \gg 1$, the argumentation can be transferred directly from the Moran process, which leads to a violation of the $1/3$-rule again. In other words, selection never favors one strategy replacing the other in coordination games with finite $w$ and $N \to \infty$ as illustrated in Fig. \ref{onethirdfig}. Therefore, the traditional ESS concept is sufficient in this limit.

\section{Fixation Times}
 \label{timesec}
 
\begin{figure}[hpbt]
\begin{center}
{\includegraphics[angle=270,width=8.3cm]{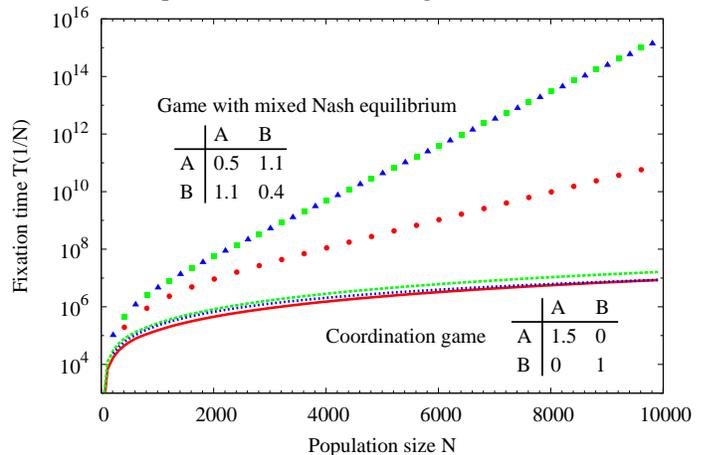}}
\caption{(color online) For a game with mixed Nash equilibrium (payoff matrix given in the figure), the conditional average fixation time for a single mutant $T(1/N)$ increases exponentially with the population size even for weak selection, $w=0.001$ (symbols). The fixation times for the local update process (filled squares) and the Wright-Fisher process (triangles) are indistinguishable (Note, however, that the intrinsic time-scales differ by a factor N, see text). The corresponding fixation times for the Moran process (circles) increase slower, but still exponentially. 
Lines: For a coordination game (payoff matrix given in the figure, $w=0.1$), the conditional average fixation time for a single mutant $T(1/N)$ decreases slower than exponentially for 
the Moran process (full line),  
the local update (dashed line), and 
the  Wright-Fisher process (dotted line). Specifically, for the Moran process 
it has been shown that it increases as $\sim N \ln N$ \cite{antal:2005aa}.}
\label{timefig}
\end{center}
\end{figure}

So far, we have only considered fixation probabilities. However, when the conditional average fixation time $T(x_0)$ for fixation in $x=1$ starting at $x_0$ becomes very large, fixation probabilities are of limited interest. For systems described by Eq.\ (\ref{langevin}), they can be computed in an elegant way
based on the Backward Kolmogorow equation \cite{ewens:1979qe}. In particular, $T(x_0)$ is given by
\begin{equation}
T(x_0) = N \int_0^1 t(x,x_0) dx,
\label{timeeq}
\end{equation}
where time is measured in elementary time steps and 
\begin{eqnarray}
t(x,x_0) &=& \frac{2 (1-\phi(x_0)) \phi(x)}{\phi(x_0) b^2(x)} e^{\int_0^x \Gamma(z) dz} S(x) 
 \nonumber
\hspace{-0.1cm} {\rm} \hspace{0.5cm} 0 \leq x \leq x_0\\ \nonumber
t(x,x_0) &=& \frac{2 \phi(x)}{b^2(x)}e^{\int_0^x \Gamma(z) dz} (S(1)-S(x)) 
\hspace{0.1cm} {\rm} \hspace{0.5cm} x_0 \leq x \leq 1.
\end{eqnarray}
For neutral selection, $w=0$, we have $S(x) = \phi(x) =x$, and $b(x) = \sqrt{2 x (1-x)/N}$ for the Moran process. 
 $T(x_0)$ is given by
\begin{eqnarray}
T(x_0) & = & - N^2\left({1}/{x_0}-1 \right) \ln ( 1- x_0).  
\end{eqnarray}
For $x_0=1/N$ and large $N$, this reduces to $T(1/N)\approx N (N-1)$, which is
identical to the result given in \cite{antal:2005aa}.
For the local update process, we obtain $T(1/N)\approx 2 N (N-1)$
for $N \gg 1$. Similarly, the Wright-Fisher process yields $b(x) = \sqrt{x(1-x)}$, which results in the time $T(1/N)\approx 2 (N-1)$.
For $w>0$, the fixation times can be computed from the integral Eq.\ (\ref{timeeq}).
As this can only be done numerically in general, two examples are given in Fig.\ (\ref{timefig}), where the asymptotics derived in \cite{antal:2005aa} for the Moran process with $w=1$ is found for all three processes. Since any $w<1$ can be mapped to $w=1$, this asymptotics is of general validity \cite{claussen:2005eh}. 

\section{Summary}

We have shown how to naturally relate the concept of ESS$_{\rm N}$ and ESS. 
In this context, the quantity $N w$, i.e.\ the intensity of selection times the population size, plays the role of an order parameter which determines the concept of evolutionary stability under consideration. 
Similarly, in population genetics the product of fitness difference and population size is the relevant parameter \cite{ewens:1979qe,crow:1970ck}.
For $N w \ll 1$ we find that the $1/3$-rule discussed by Nowak {\em et al.} \cite{nowak:2004pw} for the Moran process is generally valid and extends from the consideration of a single mutant to the more general case of the invasion of a small number of mutants. Moreover, the concept of evolutionary stability in finite populations extends to other coevolutionary processes beyond the scope of the frequency dependent Moran process.
However, for fixed $w$ and $N \gg 1$, this concept is replaced by the traditional concept of evolutionary stability. We have demonstrated that the  transition between these concepts is continuous, although the function that links both processes may be distinct for different processes.
The results for the fixation times match perfectly with this picture: Whenever selection leads to 
a pure strategy, fixation is faster than for neutral selection \cite{antal:2005aa}. 
However, in games with mixed Nash equilibria, the time for the fixation of a pure strategy increases exponentially with $N$, which is consistent with a very fast decay of the stationary distribution of the Moran process \cite{claussen:2005eh} and the prediction of the replicator dynamics that the pure strategies are unstable fixed points.

\acknowledgments{
We thank two anonymous Referees for their constructive remarks. 
A.T.\ acknowledges support by the ``Deutsche Akademie der Naturforscher Leopoldina'' (Grant No.\ BMBF-LPD 9901/8-134). 
J.M.P.\ acknowledges financial support from FCT, Portugal. 
}

\bibliographystyle{h-physrev3}

\end{document}